\documentclass[showkeys,secnumarabic,showpacs,amsmath,amssymb,prd]{revtex4}

\begin{document}

\title{A varying-c cosmology}

\author{Hossein Shojaie}
 \email{h-shojaie@sbu.ac.ir}

\author{Mehrdad Farhoudi}
 \email{m-farhoudi@sbu.ac.ir}

\affiliation{Department of Physics, Shahid Beheshti University,\\
Evin, Tehran 1983963113, Iran}

\begin{abstract}
We develop a new model for the Universe based on two key
assumptions: first, the inertial energy of the Universe is a
constant, and second, the total energy of a particle, the inertial
plus the gravitational potential energy produced by the other mass
in the Universe, is zero. This model allows the speed of light and
the total mass of the Universe to vary as functions of
cosmological time, where we assume the gravitational constant to be
a constant. By means of these assumptions, the relations between the
scale factor and the other parameters are derived. The Einstein
equation, by making it compatible with varying-$c$, is used and the
Friedmann equations in this model are obtained. Assuming the matter
content of the Universe to be perfect fluids, the model fixes
$\gamma$ to be 2/3. That is, the whole Universe always exhibits a
negative pressure. Moreover, the behavior of the scale factor is
the same for any value of the curvature. It is also shown that the
Universe began from a big bang with zero initial mass and
expands forever even with positive curvature, but it is always
decelerating. At the end, solutions to some famous problems, mainly
of the standard big bang model, and an explanation for the
observational data about the accelerating Universe are provided.



\end{abstract}

\maketitle

\section{Introduction}
The Standard Big Bang (SBB) model of the Universe is based on the
cosmological principle, the Weyl postulate, and the general
relativity~\cite{d'Inverno-1992}. It is the most successful model
for describing the structure of the Universe. The coincidence
between its predictions, for example, the abundance of the particles and
the astronomical observations has put other models, such as the
steady-state model, aside~\cite{Liddle-2003}. But, in spite of these
abilities, the SBB is not a complete theory, as its initial conditions
have weakened its strength. Some of these built-in assumptions are
known as the horizon problem, the flatness problem, and the relic-particle abundance problem~\cite{Magueijo-1999}.

To solve these problems, the idea of the inflationary Universe has been
introduced~\cite{Guth-1981,Linde-1982,Linde-1983,Kolb-Turner-1990}.
It solves almost all the problems of the SBB, but until now, no universal
acceptable microscopic foundation has been provided for this
scenario~\cite{Albrecht-Magueijo-1999}. In addition, some new
problems, for example, the quasi-flatness problem, have
arisen~\cite{Barrow-Magueijo-1998,Brandenberger-2001}.

Apart from the problems described above, the interpretation of the
recent observations, which claim that the Universe seems to be
accelerating~\cite{Perlmutter-1997,Garnavish-1998,Schmidt-1998},
shows that the SBB model has to be modified to include a
$\Lambda$ term~\cite{Wang-2000}. But, this amendment leads to
another problem, that is, the probable incompatibility between
quantum field theory and general relativity in estimating the
vacuum energy density by an order even larger than $10^{120}$.
This latter problem is called the $\Lambda$
problem~\cite{Carroll-2000}.

In the last decade, the Varying Speed-of-Light (VSL) models have
been provided to solve the SBB problems as
well~\cite{Albrecht-Magueijo-1999,Barrow-Magueijo-1998,Moffat-1993,
Clayton-Moffat-1998,Barrow-1999,Barrow-Magueijo-1999,Albrecht-1999,
Clayton-Moffat-2000,Magueijo-2000,Youm-2001,Basset-2000}. Although
it seems that these models can be viewed as alternatives to the
inflationary scenario, non constructive structures and new
problems have debarred them from acceptance until now. For a review
of these kinds of theories and their shortcomings, see, for example,
ref.~\cite{Magueijo-2003} and references therein.

One of the crucial criticism against varying-$c$ theories is that
the Lorentz invariance \emph{may} break down, therefore, the special
relativity should be adjusted. However, in the
literature~\cite{Jafari-Shariati-2003}, people claim that the
Fock--Lorentz~\cite{Fock-1964,Manida-1999} and the
Magueijo--Smolin~\cite{Magueijo-Smolin-2002,Magueijo-Smolin-2003}
transformations, which can be regarded as examples of the VSL theories,
are only re-descriptions of the special relativity, and hence, the
Lorentz invariance can be preserved. Also, according to textbooks,
for example, ref.~\cite{Rindler-1979}, besides the Euclidicity and the
isotropy assumptions, the first principle of the special relativity,
namely, the relativity principle, leads solely to either Galilean or
Lorentz transformations. In the latter transformation, there is an
upper local limit for speed of particles. The semistrong principle
of equivalence permits the possibility of different numerical
contents, namely, different values of the fundamental constants, at
different parts of space-time in the Universe. Hence, the global
laws, whose local approximations are considered in the various local
space-time regions, probably involve the derivatives of these
constants~\cite{Rindler-1979}.

Until now, there is no theory of physics that has determined the values of fundamental physical constants. Even more,
none has specified wether these constants are true constants or not.
Practically, these constants may be regarded as initial values in
theories. It is worth noting that there are two different kinds of
physical constants. The first group are constants that are related
to phenomena or quantities. The second group are those that appear
as coefficients in relations. Examples of the first group are $c$,
as the speed of light (a quantity) or $c$, as the ultimate speed of
all material particles (a phenomenon), and $e$ the electric charge
unit. On the other hand, $G$, the gravitational constant, and $H$,
the Hubble parameter, are examples of the second group. Although the
values of constants in both groups strictly depend on the choice of
units, but dimensionless ratios of constants of the first group
should rather be determined from their theories themselves.

It is plausible to assume that cosmological laws are global laws and
conventional physical laws are special (local) cases of them. Also,
these global laws would probably be simple in essence and format,
and the complexities of our present laws (if any) are perhaps caused
by limitations that localities have forced on them. An optimistic
and simple insight is that, generalizations of the physical
laws of today may lead to cosmological laws. Implicitly, as the Universe can
be regarded as a collection of statistical quantities, a numerical
analysis of the ratios of these quantities would be meaningful even though there is a lack of any reasonable theory to explain
them, for example, the large-numbers hypothesis~\cite{Dirac-1937}, though
some people believe that the Dirac large-numbers hypothesis has some
cracks in. Therefore, it seems acceptable that a cosmological model
depends more on symmetries than on initial conditions.

Inspired by the above discussions and some postulates presented in
ref.~\cite{Albrecht-Magueijo-1999}, a simple varying-$c$
cosmological model is provided in this work, though it 
differs somehow from the former ones in that it does not solve
cosmological problems only. That is, it is based on a few
cosmologically plausible assumptions through inertial arguments, to
be described in the next section, in addition to the assumption that
the speed of light is not a constant. Other differences will be
obvious through the work and also, we emphasis a few key
distinctions in the conclusion. Although, the authors never claim
that it is a complete model, to show its abilities, solutions to
some typical problems are given.

In Section~\ref{Basic0}, basic concepts of our model are introduced.
Two fundamental assumptions and the concepts behind them are determined,
and then variations of the essential quantities with respect to the
scale factor are derived. In Section \ref{Field0}, the
Robertson-Walker metric and the Einstein equation are discussed and
hence the Friedman and field equations are obtained. Also, the time
dependency of quantities in the former section are derived. The age
of the Universe is calculated in terms of the Hubble constant.
Section~\ref{Problem0} provides solutions to some famous problems
such as the horizon, the accelerating Universe, the flatness, and the
quasi-flatness problems.

\section{Basic concepts}\label{Basic0}

Before proceeding, let us specify the frame work and the metric
that we are going to use. The cosmological principle, i.e., the
assumptions of homogeneity and isotropy of the Universe, implies
that there should be no preferred point in the Universe. This is
obtained when the spatial part of the metric has a constant
curvature. The Robertson-Walker metric is the most general metric
that merits these properties even if $c_0$ is replaced by $c(t)$,
i.e.,
\begin{equation}
ds^2=c(t)^2dt^2-a(t)^2\left(\frac{dr^2}{1-kr^2}+r^2d\Omega^2\right)\
.\label{Field29}
\end{equation}
Rewriting~(\ref{Field29}) as
\begin{equation}
ds^2=\left(\frac{c(t)}{c_0}\right)^2\left[c_0^2dt^2-a'(t)^2
\left(\frac{dr^2}{1-kr^2}+r^2d\Omega^2\right)\right]\
,\label{Field30}
\end{equation}
shows that the metric~(\ref{Field29}) is conformal to the
original one, i.e., the metric~(\ref{Field29}) where $c(t)$ is
replaced by $c_0$. That is, the causal structures of these metrics
are the same. The components of this metric are
\begin{equation}
g_{00}=1\ ,\quad g_{11}=-\frac{a(t)^2}{1-kr^2}\ ,\quad
g_{22}=-a(t)^2r^2\ , and \quad g_{33}=-a(t)^2r^2\sin^2\theta \
.\label{Field33}
\end{equation}
Note that $g_{00}$ is the coefficient of
$dx^0\left(=c(t)dt\right)$ and not $dt$. (In a similar way, one
can introduce another coefficient, $c'(t)$, in the relation
$x^0=c'(t)t$, where in this model due to the later
relation~(\ref{Field23}), $c'(t)=\frac{5}{4}c(t)$.) This
guarantees that all functions (including the components of the
tensors in the general relativity) derived from this metric are
identical with those derived from the usual metric with
constant-$c$ but now with $dx^0$ instead of $dt$. The only task to
do, when someone wants to show the time dependency of the
functions, is to replace $\partial x^0$ by $c(t)\partial t$ in
the relations. In the rest of this manuscript, all arguments are hold
in the Robertson-Walker framework as the comoving (preferred)
frame of cosmology.

On the other hand, one may also prefer to write~(\ref{Field29}) in
another coordinate, such as $c(t)^2dt^2=c_0^2dt'^2$, and
hence to get
\begin{equation}
ds^2=c_0^2dt'^2-a(t')^2\left(\frac{dr^2}{1-kr^2}+r^2d\Omega^2\right)\
.\label{Field35}
\end{equation}
That is, in terms of this new time, $t'$, the metric is reduced to
the standard Robertson-Walker metric. This may bring ambiguities
about which metric is physical. As mentioned before, the
metric~(\ref{Field29}) is conformal with the standard
Robertson-Walker metric and so one cannot distinguish which metric
is the real physical one.

In a cosmological model, it is constructive to specify whether the
mass-energy conservation law holds or not. Although some few models
are based upon the idea of changing-$M$, for example, the steady-state
theory, in most theories, it has been accepted that the whole
mass $M$ of the (observable) Universe is a constant. While the speed
of light is supposed to be a constant, there is no difference
between the constancy of the total mass $M$ and the total energy
$Mc^2$ of the (observable) Universe, but with varying-$c$, for example, in
the VSL theories, one should strictly decide about the constancy of
the mass and the energy content of the Universe separately. Hence,
in the preferred frame of the cosmology, we begin with the following
two key assumptions that are related directly to the above facts

\begin{description}

    \item[Assumption(1):] The inertial energy of the (observable) Universe,
    measured in the preferred frame of cosmology, is a constant, i.e.,

    \begin{equation}
    Mc^2=const\ ,\label{Basic2}
    \end{equation}
    i.e., the energy conservation law is valid.

    \item[Assumption(2):] The total energy of a particle with mass $m$, including its inertial energy
    and its gravitational potential energy produced by the mass of
    the (observable) Universe, measured in the preferred frame of cosmology, is zero, i.e.,

    \begin{equation}
    -\frac{GMm}{R}+mc^2=0\ ,\label{Basic1}
    \end{equation}
    where $R$ is the radius of the (observable) Universe\footnote{For relativistic massless particles , one should simply
    replace $mc^2$ by their energies and $m$ by their corresponding mass, $E/c^2$.}. In other words,
    the inertial energy of a particle is due to
    the gravitation potential energy of the mass content of
    the Universe upon it.

\end{description}

These two assumptions have old analogous backgrounds, for the first
one, it is what has been accepted in most theories, however,
here without a limitation rule that force the speed of light, $c$,
to be a constant. Hence, in general, both $M$ and $c$ can be assumed
to be functions of coordinates, in addition, to the homogeneity
assumption, in the preferred frame of the cosmology, it is plausible
to assume that $M$ and $c$ depend on the cosmological time only. The
second relation, which points out that the inertial energy
of a particle is compensated for by the negative gravitational potential
energy due to the interaction with all matter of the Universe, is
indeed a mathematical formulation of the Machian view about the
inertial energy of a particle. Hence, particles may be created
without violation of the energy conservation law. Also it should be noted
that, in~(\ref{Basic1}), $m$ can be canceled out, and
hence one can write this relation in the cosmological preferred
frame instead of going to the rest frame of the particle.

One may feel that the foundation of these assumptions is unsatisfactory,
for one may argue that these are notions of weak-field gravity such
as the Newtonian potential energy used in the context of the large-scale gravitational field. However, one should note that even the
Friedmann equations can be obtained from the Newtonian approach.
Besides, constructing an energy quantity by dimensional analysis
using cosmological parameters such as $M$, $R$, and $G$, one can
gain the Newtonian potential energy in its simplest form. Moreover,
to proceed experimental data, one should use these usual notions.
Indeed, these justifications are not the first goal of researchers into varying-$c$ models, but its
consequences. In other words, this is a simple model only, and its
subsequent results and discussions will prove it.

Although one can think of $G$ as a variable, a simple case for a
global theory may be provided with $G$ independent of time.
In this work, the authors assume it to be a constant, but to have $M$, the total
mass of the Universe, as a varying parameter. That is, the probable variation of $G$ that
appeared in the other models is ascribed to varying $M$ in this model.  In this sense, the Newtonian
gravitational potential for a fixed-mass particle at a fixed
distance does not depend on (cosmological) time.

Adding up~(\ref{Basic1}) for all particles in the
Universe, one gets
\begin{equation}
-\frac{GM\sum m}{R}+\sum mc^2=0\ ,
\end{equation}
where one can substitute $\sum m$ with $M$. Consequently it gives
\begin{equation}
Mc^2=\frac{GM^2}{R}\ ,\label{Basic3}
\end{equation}
which determines the value of $Mc^2$ in~(\ref{Basic2}). It also reduces to
\begin{equation}
\frac{GM}{Rc^2}=1\ .\label{Basic19}
\end{equation}
Obviously, as mentioned earlier, canceling $m$ in~(\ref{Basic1}) gives~(\ref{Basic3})
and~(\ref{Basic19}). The order of the last relation can be obtained separately
by constructing a dimensionless relation from the
fundamental cosmological parameters, as mentioned in
the famous paper by Brans-Dicke~\cite{Brans-Dicke-1961} and with a
somewhat different notation in Sciama's paper~\cite{Sciama-1953}.
By considering~(\ref{Basic19}) and following some
authors who use $\frac{GM}{Rc^2}$ for comparing the gravity
intensities of the celestial objects, one can think of the
Universe as twice as strong as black holes. This will be
discussed in more details in a later work. Rearranging~(\ref{Basic19}) results in
\begin{equation}
\frac{GMc^2}{R}=c^4\ .\label{Basic4}
\end{equation}
The numerator of the left-hand side of~(\ref{Basic4}) is
a constant by our assumptions, hence differentiating~(\ref{Basic4})
with respect to the cosmological time, $t$, gives
\begin{equation}
-\frac{\dot R}{R}=4\frac{\dot c}{c}\ .\label{Basic5}
\end{equation}
If one now defines the scale factor $a$ as usual, and assumes that
$R\propto a$,~(\ref{Basic5}) leads to
\begin{equation}
\frac{dc}{c}=-\frac{1}{4}\frac{da}{a}\ .\label{Basic6}
\end{equation}
Integrating~(\ref{Basic6}) yields
\begin{equation}
c=c_0\left(\frac{a_0}{a}\right)^\frac{1}{4}\ ,\label{Basic7}
\end{equation}
where by $c_0$ and $a_0$, we mean the current values of these
quantities. Defining the redshift, $1+z\equiv\frac{a_0}{a_z}$ as
usual, one finds from~(\ref{Basic7}) that
\begin{equation}
\frac{c_z}{c_0}=\left(1+z\right)^\frac{1}{4}\ .\label{Basic13}
\end{equation}

Following the same procedure used in deriving~(\ref{Basic7}), one can get
\begin{equation}
M=M_0\left(\frac{a}{a_0}\right)^\frac{1}{2}\ .\label{Basic8}
\end{equation}

To determine the relationship between the wavelength $\lambda$
and the scale factor $a$, using the usual method, one has for two
light rays with a time interval $dt$ between two points that
\begin{equation}
\frac{c(t_e)dt_e}{a(t_e)}=\frac{c(t_r)dt_r}{a(t_r)}\ ,
\end{equation}
where $t_e$ and $t_r$ are the time of emission and reception,
respectively. Substituting $\lambda=cdt$ in the above relation yields
\begin{equation}
\lambda=\lambda_0\frac{a}{a_0}\ ,\label{Basic11}
\end{equation}
as usual. Another way to derive a relation for $\lambda$ is to
suppose a stationary wave between two points. For this wave to
remain stationary during the evolution of a non static Universe, one gains again
the relation~(\ref{Basic11}). Using $f=c/\lambda$ for the
frequency, and~(\ref{Basic7}) and~(\ref{Basic11}),
implies that
\begin{equation}
f\propto a^{-\frac{5}{4}}\ .\label{Basic12}
\end{equation}

A question arises when one compares~(\ref{Basic11})
and~(\ref{Basic12}): which one corresponds to the redshift that
one detects, the wavelength or the frequency? The answer is the
frequency, for all the astronomical observation tools are
sensitive to frequency, i.e., the number of the peaks that one
receives in a unit time. Naturally, knowing the speed of light,
one can claim that one measures the wavelength of the light
simultaneously. Hence, a useful relation for the redshift is
\begin{equation}
\frac{f_e}{f_r}=\left(\frac{a_0}{a_z}\right)^\frac{5}{4}=\left(1+z\right)^\frac{5}{4}\
,\label{Basic20}
\end{equation}
where $f_e$ and $f_r$ are the frequency at the time of emission
where $a=a_z$, and the time of reception where $a=a_0$,
respectively. This relation shows that the frequency displacement
in the spectrum of a cosmological object corresponds to a
newer epoch than the other models expect, for example, for
$\frac{f_e}{f_r}=2$, the relations implies that the corresponding
object lies at~$z=0.74$ when the Universe had a radius $57\%$
smaller than it is nowadays, while other theories
predict these values as $z=1$ and $\frac{a_z}{a_0}=0.5$. It is
evident that for a nearer object, this effect is more negligible.

Now let us have a look at the energy density distribution of the
black-body radiation, i.e.,
\begin{equation}
\epsilon(f)=8\pi
h\left(\frac{f}{c}\right)^3\frac{1}{\exp\left(\frac{hf}{kT}\right)-1}\
.\label{Basic17}
\end{equation}
The left-hand side of~(\ref{Basic17}), the energy density,
should vary as $a^{-3}$, and so the right-hand side should do the
same. The fraction $(f/c)^3$ on the right-hand side, which
by~(\ref{Basic7}) and~(\ref{Basic12}) evolves proportionally with
$a^{-3}$, guarantees this condition. The simplest choice for the
exponential function in the denominator to preserve the behavior
of the distribution is to assume
\begin{equation}
T\propto a^{-\frac{5}{4}}\ ,\label{Basic16}
\end{equation}
or consequently
\begin{equation}
T=T_0\left(\frac{a_0}{a}\right)^\frac{5}{4}\ .\label{Basic21}
\end{equation}
In deriving~(\ref{Basic16}), we have assumed that the
Planck's constant, $\hbar$, and the Boltzmann constant, $k_B$,
plausibly to be constants.

Also some calculations using~(\ref{Basic2}) and~(\ref{Basic8}) imply that
\begin{equation}
\rho\propto \frac{M}{R^3}\propto \frac{1}{a^\frac{5}{2}}\
.\label{Basic14}
\end{equation}
So one can write the relationship between $\rho$ and $a$ as
\begin{equation}
\rho=\rho_0\left(\frac{a_0}{a}\right)^\frac{5}{2}\ ,\label{Basic15}
\end{equation}
in spite of the matter content and any epoch of the Universe.
Here, only the gravitational effects of the density $\rho$ are
considered. In general, the density  includes the ordinary matter, the
radiation, dark matter, dark energy, and any other forms of
particles. For those who are uncomfortable with this result, one
can write the familiar form
\begin{equation}
\rho c^2=\rho_0{c_0}^2\left(\frac{a_0}{a}\right)^3\
,\label{Basic18}
\end{equation}
which is provided for the matter-dominated epoch of the Universe
(i.e., dust) in the famous models. However, one should remember
that in this situation, regardless of the types of the content of
the Universe, it (or at least the dominated part of it) has to
obey~(\ref{Basic15}) and~(\ref{Basic18}).

The introduction of varying-$c$ forces one to rethink some of the
standard lore of cosmology. As this model does not allow an
earlier radiation-dominated stage of the Universe, it may be
better to clear up how the observed Cosmic Microwave Background (CMB)
would arise, or to what extent nucleosynthesis would be allowed to
occur. However, at first glance, one should note that the origin
of CMB was when electrons were bounded in atoms. It can occur
regardless of the stage of the Universe. The nucleosynthesis seems
to be rearranged.

\section{The field equation}\label{Field0}

The Einstein field equation is a well-constructed theory that can
be considered even if the fundamental constants are not really
constants, i.e. the relation $G_{ab}\propto T_{ab}$ generally
implies that geometry and matter affect each other
(independent of the frame of reference). Due to the correspondence
principle, one can derive the full equation as
\begin{equation}
G_{ab}=\frac{8\pi G}{c^4}T_{ab}\
.\label{Field12}
\end{equation}
Also the definition of the metric components~(\ref{Field33}) is
compatible with the Bianchi identity, i.e.,
\begin{equation}
\nabla_aG^{ab}=0\ .\label{Field34}
\end{equation}

Assuming a varying-$c$ model and the correspondence principle, if
one tries to use the Einstein equation as a field equation in the
cosmological frame, at least some points should be noted. Relation~(\ref{Basic7}) implies that the coefficient of the
Einstein equation is also varying with time, i.e.,
\begin{equation}
\frac{\dot \kappa}{\kappa}=\frac{\dot a}{a}\quad \textrm{or}\quad
\kappa\propto a\ ,\label{Field1}
\end{equation}
where $\kappa=8\pi G/c^4$. This means that the coefficient of the
Einstein equation varies proportionally to the scale factor, so the coupling between
geometry and matter, becomes stronger as the Universe
expands. Conversely, for the case when $a\rightarrow 0$ obviously
$\kappa\rightarrow 0$. By applying the covariant divergence to both sides of the Einstein equation using~(\ref{Field34}), it turns
out that
\begin{equation}
\nabla_a\left(\frac{8\pi G}{c^4}T^{ab}\right)=0 \ ,\label{Field2}
\end{equation}
and hence, one can not get the canonical energy-momentum
conservation relation, $\nabla_aT^{ab}=0$. A crucial task in this
approach remains to provide a Lagrangian formulation for~(\ref{Field12}), though, it is obvious that such a
variation must provide the whole argument of~(\ref{Field2}), i.e., $\frac{1}{c^4}T^{ab}$. In other
words, the standard definition of the energy-momentum tensor needs
to be modified to consider variations in $c$, for example, a change in
$c$ should act as a source of matter. To keep the usual
definition of the energy-momentum tensor, one may add other
term(s) to $T^{ab}$. This is the aim of another work of the
authors~\cite{Shojaie-Farhoudi-2004-2}.

As one works with the Robertson-Walker metric in the cosmological
frame, where $c$ is assumed to be a function of the cosmological time
only,~(\ref{Field2}) yields
\begin{equation}
\partial_a\left(\frac{1}{c^4}\right)T^{ab}
+\frac{1}{c^4}\nabla_aT^{ab}=0\ .\label{Field3}
\end{equation}
Assuming $T_{ab}$ to be a perfect fluid, i.e.,
\begin{equation}
T_{ab}=\left(\rho+\frac{p}{c^2}\right)u_au_b-pg_{ab}\ ,
\end{equation}
where $u_a=(c,0,0,0)$ in the co-moving frame, the spatial
equations of~(\ref{Field3}) are identically zero. The time equation derived
from~(\ref{Field3}) by using $p=(\gamma-1)\rho c^2$, reduces to
\begin{equation}
\frac{\dot\phi}{\phi}+3\gamma\frac{\dot a}{a}
+\frac{\dot\epsilon}{\epsilon}=0\ .\label{Field4}
\end{equation}
where $\phi=\frac{1}{c^4}$ and $\epsilon=\rho c^2$. Equation
(\ref{Field4}) can be treated as the field equation. Simplifying
the above equation using the relation~(\ref{Basic7}) leads to
\begin{equation}
\rho\propto \frac{1}{a^{3\gamma+\frac{1}{2}}}\quad
\textrm{and}\quad \rho c^2\propto \frac{1}{a^{3\gamma+1}}\
.\label{Field5}
\end{equation}
Comparing~(\ref{Field5}) and~(\ref{Basic15}) with
each other, yields
\begin{equation}
\gamma=\frac{2}{3}\ ,\label{Field14}
\end{equation}
i.e., a fixed $\gamma$ that always gives $p=-\frac{1}{3}\rho
c^2$. Therefore, the Universe has a negative pressure all the time,
and there is no such dust- or radiation-dominated epochs.

Using the Robertson-Walker metric in the Einstein equation and
substituting $c\partial t$ instead of $\partial x^0$ leads to an analogy of the
Friedman equations, i.e.,
\begin{equation}
\left(\frac{\dot a}{a}\right)^2+\frac{kc^2}{a^2}=\frac{8\pi
G}{3}\rho\ ,\label{Field10}
\end{equation}
and
\begin{equation}
2\frac{\ddot a}{a}-2\frac{\dot c}{c}\frac{\dot
a}{a}+\left(\frac{\dot a}{a}\right)^2+\frac{kc^2}{a^2}=-\frac{8\pi
G}{c^2}p\ .\label{Field11}
\end{equation}
As can be easily seen, the term $-2\frac{\dot c}{c}\frac{\dot
a}{a}$ is the only difference between these equations in this case
and the usual classical ones. One can also check that combining
these two relations and removing the $\ddot a$ phrase gives~(\ref{Field5}) again. By
substituting~(\ref{Field10}) in the latter, one gains
\begin{equation}
\frac{\ddot a}{a}-\frac{\dot c}{c}\frac{\dot a}{a}=-\frac{4\pi
G}{3}\left(\rho+\frac{3p}{c^2}\right)\ .\label{Field15}
\end{equation}
Using~(\ref{Basic7}) and $p=(\gamma-1)\rho c^2$ in
the last relation, it yields
\begin{equation}
\frac{\ddot a}{a}=-4\pi
G\left(\gamma-\frac{2}{3}\right)\rho-\frac{1}{4}\left(\frac{\dot
a}{a}\right)^2 \ .\label{Field16}
\end{equation}
Now by applying~(\ref{Field14}) in the above, one can write the
second Friedmann equation simply as
\begin{equation}
\frac{\ddot a}{a}=-\frac{1}{4}\left(\frac{\dot a}{a}\right)^2 \
.\label{Field24}
\end{equation}
We will return to this equation later.

Now, to solve the Friedmann equations, one should substitute~(\ref{Basic7}) and~(\ref{Basic15}) in~(\ref{Field10}), hence it gives
\begin{equation}
\left(\frac{\dot a}{a}\right)^2=\frac{8\pi
G}{3}\rho_0\left(\frac{a_0}{a}\right)^\frac{5}{2}
-\frac{kc_0^2}{a_0^2}\left(\frac{a_0}{a}\right)^\frac{5}{2}
=H_0^2\left(\frac{a_0}{a}\right)^\frac{5}{2}\ ,\label{Field13}
\end{equation}
where $H_0$ is the value of the Hubble parameter
$H(t)\equiv\frac{\dot{a}}{a}$ at the current time, $t_0$. Taking the
square root of the above relation and noting that the
negative sign has no meaning, (for in this case it gives $H_0=-H_0$!), one has
\begin{equation}
a^\frac{1}{4}da=H_0a_0^\frac{5}{4}dt\ .\label{Field18}
\end{equation}
Integrates~(\ref{Field18}) and setting $a=0$ for $t=0$, i.e., the Universe
has begun from a big bang, one gets
\begin{equation}
a=a_0\left(\frac{t}{t_0}\right)^\frac{4}{5}\ ,\label{Field19}
\end{equation}
where
\begin{equation}
t_0=\frac{4}{5H_0}\ .\label{Field20}
\end{equation}
It is worth noting that the differential equation~(\ref{Field18}), because of the plausible assumption that $a\geq0$, always leads to a big bang.

From the best measurements, the Hubble constant is
\begin{equation}
H_0=100h\frac{km}{s}\frac{1}{Mpc}\quad \textrm{where}\quad
0.55\leq h\leq 0.8\ .\label{Field21}
\end{equation}
One should note that the methods of measuring $H_0$
may be affected by varying-$c$ in this model, for example, corrections in
measuring the redshift as mentioned before. Hence, the value of the
Hubble constant may differ from its predicted range. In spite of
the above argument, using the data of~(\ref{Field21})
in~(\ref{Field20}) gives the age of the Universe to as
\begin{equation}
9.78 \times 10^9\ years < t_0 < 14.22 \times 10^9\ years\
\label{Field22}
\end{equation}
which is a better approximation than the current values from
the most accepted models. Note that~(\ref{Field13})
implies that $\dot a$ never changes sign even if $k\neq 0$,
for $\dot a$ to become zero, $a$ should tend to infinity. The only
acceptable choice is $\dot a>0$, i.e., the Universe expands forever and
\begin{equation}
\lim_{t\rightarrow \infty}\dot a=0\ ,\label{Field17}
\end{equation}
in spite of the sign of $k$. In agreement with the above conclusion,
one can rewrite~(\ref{Field24}) as
\begin{equation}
\frac{\ddot a}{a}=-\frac{1}{4}\left(\frac{\dot
a}{a}\right)^2=-\frac{1}{4}H^2\ .\label{Field26}
\end{equation}
The above relation indicates that the Universe always decelerates
during its life independent of its topology and that this
deceleration rate can be explained completely by the Hubble parameter.
Note that even with $k=+1$, one gets an ever-expanding Universe
with a closed topology. As it can obviously be seen, the deceleration parameter
$q\equiv-\frac{\ddot aa}{\dot a^2}$, is always
\begin{equation}
q=\frac{1}{4}\ .
\end{equation}
From~(\ref{Field26}), one gets
\begin{equation}
\lim_{t\rightarrow \infty}\ddot a=0\ .\label{Field27}
\end{equation}
Hence,~(\ref{Field17}) and~(\ref{Field27}) imply
that no \emph{Big Crunch} will be visited even with a positive curved
space-time.

The relations~(\ref{Basic7}) and~(\ref{Field19}) give
\begin{equation}
c(t)=c_0\left(\frac{t_0}{t}\right)^\frac{1}{5}\ ,\label{Field23}
\end{equation}
with the limits
\begin{equation}
\lim_{t\rightarrow0}c=\infty\quad\textrm{and}\quad\lim_{t\rightarrow\infty}c=0\
.\label{Field31}
\end{equation}
Also~(\ref{Basic15}) and~(\ref{Field19}) yield
\begin{equation}
\rho(t)=\rho_0\left(\frac{t_0}{t}\right)^2\ ,\label{Field25}
\end{equation}
where one has
\begin{equation}
\lim_{t\rightarrow0}\rho=\infty\quad\textrm{and}\quad\lim_{t\rightarrow\infty}\rho=0\
,\label{Field32}
\end{equation}
which implies at $t=0$, one still meets the singularity. Also combining~(\ref{Basic21}) and~(\ref{Field19}) yields
\begin{equation}
T=T_0\frac{t_0}{t}\ . \label{Field37}
\end{equation}
Therefore,
\begin{equation}
\lim_{t\rightarrow0}T=\infty ,\label{Field7}
\end{equation}
i.e., this model is also a hot big bang model.

\section{Solving some problems}\label{Problem0}

One of the most important reason for defining a new cosmological
model is to be able to solve the problems of the other models. In
this section, we are going to provide some solutions to the problems
mainly encountered with SBB, for the model introduced in this work
has a big bang as well.

\subsection{The horizon problem}

The horizon problem implies that because of the limited speed of
light, all the regions of (at least) the (observed) Universe can
not be in contact until the time of observation. So the different
parts of the Universe may behave in different ways. The
observations disagree with this idea. To solve this problem in
this model, one should notice that the speed of light had no upper limit on it, i.e.,
\begin{equation}
c\rightarrow\infty\quad\textrm{as}\quad t\rightarrow0\
.\label{Horizon1}
\end{equation}
Mathematically speaking, one can calculate the particle horizon
\begin{equation}
\int^{r_0}_0\frac{dr}{\sqrt{1-kr^2}}=\int^{t_0}_0\frac{c(t)dt}{a(t)}=
\int^{t_0}_0\frac{c_0t_0}{a_0}\frac{dt}{t}=\frac{c_0t_0}{a_0}\ln
t\Big|^{t_0}_0\ ,\label{Horizon2}
\end{equation}
which is clearly infinite, i.e., all the regions of the Universe
are observable. As one can claim that this argument is valid for
any position and any time in the Universe, this means that all the
regions of the Universe are causally connected.

\subsection{The flatness and quasi-flatness problem}

The first Friedmann equation~(\ref{Field10}) shows that in this
model the critical density, $\rho_c$, which corresponds to the
flat space-time, is as usual
\begin{equation}
\rho_c=\frac{3H^2}{8\pi G}\ .\label{Flat4}
\end{equation}
Defining $\Omega\equiv\frac{\rho}{\rho_c}$ and rewriting~(\ref{Field10}) for $k\neq0$ gives
\begin{equation}
\left|\Omega-1\right|=\frac{c^2}{a^2H^2}\ ,\label{Flat1}
\end{equation}
which implies that in the SBB model the right-hand side of the above
equation is proportional to the positive power law of $t$.
Consequently, for $\Omega$ to be nearly flat, which observations
confirm, it should be finely tuned to the one in the very early
epoch of the Universe. This is known as the flatness problem.

However, in this model, using $c\propto t^{-\frac{1}{5}}$, $a\propto
t^\frac{4}{5}$ and $H\propto t^{-1}$, one has
\begin{equation}
\left|\Omega-1\right|\propto1\ .\label{Flat2}
\end{equation}
This means that $\Omega$ is not time-dependent at all and its
value does not vary during the life of the Universe. Also, the last
relation does not force the geometry to be exactly flat, which
solves the quasi-flatness problem as well. This problem arises as
the Universe does not seem to be perfectly flat. It will be shown that
the most probable value of $\Omega$ in this model is less than
unity~\cite{Shojaie-Farhoudi-2004-2}.

To see the effect of the curvature, $k$, on the
acceleration, $\ddot a$, substitute~(\ref{Field10})
into~(\ref{Field24}), it gives
\begin{equation}
\frac{\ddot a}{a}=-\frac{2}{3}\pi
G\rho+\frac{1}{4}\frac{kc^2}{a^2}\
.\label{Flat3}
\end{equation}
The first term on the right-hand side of the above equation is a negative term in power of $t^{-2}$. The second term is also a term in the power of $t^{-2}$, whose sign depends on $k$. Hence, surprisingly, when $k$ is positive the Universe decelerates with a rate less than that for the flat case, and conversely the rate of deceleration is more than that of the flat case when $k$ is negative. As can easily be checked, this special property arises from varying-$c$, i.e., the extra term $\frac{\dot c}{c}\frac{\dot a}{a}$ in (\ref{Field11}) holds the Universe near flat geometry.

\subsection{Curvature domination problem}

In SBB, one may worry about the curvature domination, i.e., for a
large enough $a$, the curvature term in~(\ref{Field10}) dominates.
In this model, as discussed above, the curvature and the density
terms have the same order of $t$ and $a$, hence, if the curvature
could not dominate in an epoch, obviously it cannot do later.

\subsection{Relic-particle problem}

The density of the massive particles decreases as
\begin{equation}
\rho_{massive}\propto\frac{1}{a^3}\ .\label{Relic1}
\end{equation}
Hence, in the early epoch of the Universe, these massive particles
should have dominated over the radiation in the SBB model. Nowadays,
we have no evidence of observing these massive particles. This is
known as the relic-particle problem and sometimes is referred to as
the magnetic monopole problem, as one of the first known massive
particles was the magnetic monopole. However, in this model the
density of the Universe, including any combinations of different
kinds of matter and (or) energy, decreases with a lower power of $a$ as
\begin{equation}
\rho\propto\frac{1}{a^\frac{5}{2}}\ .\label{Relic2}
\end{equation}
This shows that the above problem does not occur in this model.

\subsection{The accelerating Universe}

Recently, analysis of the luminosity of the supernovae at redshift
$z\sim1$ shows that they are fainter than
expected~\cite{Perlmutter-1997,Garnavish-1998,Schmidt-1998}.
Hence, it has been suggested that the Universe may experience a
period of acceleration. In our model, the Universe always
decelerates. So, how can one explain this situation?

The key solution is located in the dependency of the flux density,
$S$, of the celestial objects to the redshift, $z$. That is, the
results of these observational data would be due to different
relations for the redshift of objects. Despite the SBB model where
$S\propto(1+z)^{-2}$, in this model, one has
\begin{equation}
S\propto (1+z)^{-\frac{5}{2}}\ ,\label{Accelerate1}
\end{equation}
for the photons arriving from those objects and they have frequencies (and
obviously energies) proportional to $(1+z)^{-\frac{5}{4}}$, see~(\ref{Basic20}). Hence, for a particular redshift $z$, a
cosmological object in this model has $\frac{1}{\sqrt{1+z}}$ times
the flux density calculated by the SBB model, for example, the supernovae with
$z=1$ should be found to be fainter by a factor of approximately $29\%$.
As a result, this analyzing should be revised and we expect there
should be no need to assume an accelerating Universe.

\subsection{The source of inertia}

Relation~(\ref{Basic8}) implies that
\begin{equation}
\lim _{a\rightarrow 0}M=0\ .\label{Source1}
\end{equation}
It means that the Universe began from a big bang, $a=0$, when it was
empty. By assuming~(\ref{Basic2}), the decreasing-$c$ makes
the total mass of the Universe increase. Hence, one may
speculate that the light should act as a source of inertia due to
its reduction in speed. We are not going to provide a mechanism for
this phenomenon, however, the idea of matter creation has been
employed in other cosmological models, for example Hoyle-Narlikar and
steady-state models~\cite{Narlikar-2002}, but with a constant $c$.

\section{Conclusion}

We have constructed a new model for the Universe with two key assumptions
written in the preferred frame of cosmology: (\textsl{i}) inertial energy
of the Universe, $Mc^2$, is a constant, and (\textsl{ii}) the total energy of a
particle, i.e., its inertial energy plus its gravitational potential
energy, is zero. The second assumption is another version of the
Machian view. In addition, we do not restrict the speed of light to
a constant, however, we assume that $G$ is a constant. These and
some other plausible assumptions lead to the relations between the
scale factor and the speed of light, the total mass and the density
of the Universe. Also the relations between the wavelength and the
frequency of light and the scale factor are obtained. This suggests
that corrections to the redshift measurements should be made.

The Robertson-Walker metric and the Einstein equation compatible
with varying-$c$ are used. However, the result is
$\nabla_aT^{ab}\neq0$, which with the assumption of perfect fluids
it consequently fixes $\gamma$ at $2/3$. This means that there are
no dust- or radiation-dominated epochs, and $p=-\frac{1}{3}\rho c^2$.

The generalized Friedmann equations yield the following properties
for the Universe in this model.
\begin{itemize}
    \item It began from a hot big bang when it was empty, i.e., $M=0$
    when $t=0$.
    \item It expands forever regardless of the sign of the curvature.
    \item It always decelerates. The deceleration parameter is fixed to
    $\frac{1}{4}$.
    \item Its age lies approximately in the range $10-14\times10^{9}$ years.
    \item The speed of light varies with the cosmological time as
    $c\propto t^{-\frac{1}{5}}$.
    \item The total mass, the density and the temperature of the
    Universe vary with the cosmological time as $M\propto
    t^{\frac{2}{5}}$, $\rho\propto t^{-2}$ and $T\propto t^{-1}$,
    respectively.
\end{itemize}

The horizon, the flatness, the curvature domination, and the relic-particle problems of the standard big-bang (SBB) are solved. Besides, the flux density of
celestial objects in this model is lower than that derived from
the SBB model, i.e., these objects look fainter than the SBB expectation.

In addition, a reduction of $c$ can be contemplated as a source of
matter creation. That is, in this model, in contrast to some other
models, for example Ref.~\cite{Albrecht-Magueijo-1999}, the geometry of the
Universe is affected by varying-$c$.

\end{document}